\documentclass{article}
\usepackage{float}
\usepackage{amsmath}
\usepackage{booktabs} 
\usepackage{multirow} 
\usepackage{adjustbox}
\usepackage{PRIMEarxiv}
\usepackage{algorithm}
\usepackage{algorithmic}
\usepackage[utf8]{inputenc} 
\usepackage[T1]{fontenc}    
\usepackage{url}            
\usepackage{booktabs}       
\usepackage{amsfonts}       
\usepackage{nicefrac}       
\usepackage{microtype}      
\usepackage{fancyhdr}       
\usepackage{graphicx}       
\graphicspath{{media/}}     
\usepackage{subfigure}
\usepackage{atbegshi}

\title{Deep reproductive feature generation framework for the diagnosis of Covid-19 and viral pneumonia using chest X-ray images}

\newcommand\blfootnote[1]{%
  \begingroup
  \renewcommand\thefootnote{}\footnote{#1}%
  \addtocounter{footnote}{-1}%
  \endgroup
}

\author{
  \large
  Ceyhun Efe Kayan, Talha Enes Koksal, Arda Sevinc, Abdurrahman Gumus*\\
  Department of Electrical and Electronics Engineering \\
  Izmir Institute of Technology
}

\begin{document}
\maketitle

\begin{abstract}
The rapid and accurate detection of COVID-19 cases is critical for timely treatment and preventing the spread of the disease. In this study, a two-stage feature extraction framework using eight state-of-the-art pre-trained deep Convolutional Neural Networks (CNNs) and an autoencoder is proposed to determine the health conditions of patients (COVID-19, Normal, Viral Pneumonia) based on chest X-rays. The X-ray scans are divided into four equally sized sections and analyzed by deep pre-trained CNNs. Subsequently, an autoencoder with three hidden layers is trained to extract reproductive features from the concatenated ouput of CNNs. To evaluate the performance of the proposed framework, three different classifiers, which are single-layer perceptron (SLP), multi-layer perceptron (MLP), and support vector machine (SVM) are used. Furthermore, the deep CNN architectures are used to create benchmark models and trained on the same dataset for comparision. The proposed framework outperforms other frameworks wih pre-trained feature extractors in binary classification and shows competitive results in three-class classification. The proposed methodology is task-independent and suitable for addressing various problems. The results show that the discriminative features are a subset of the reproductive features, suggesting that extracting task-independent features is superior to the extraction only task-based features. The flexibility and task-independence of the reproductive features make the conceptive information approach more favorable. The proposed methodology is novel and shows promising results for analyzing medical image data. \blfootnote{* Corresponding author: abdurrahmangumus@iyte.edu.tr \\ \\ Machine Intelligence Research and Applications Laboratory (MIRALAB) \\  Preprint submitted. Under review.}
\end{abstract}


\maketitle
\section{Introduction}
\label{sec:intro}
The Good Old Fashioned Artificial Intelligence (GOFAI) was the earliest stage of the artificial intelligence era which adopted the rule-based approach for solving tasks \cite{russell2010artificial}. The rule-based approach is a deterministic methodology which is capable of operating with base-level information for dealing with simple problems. The rule-based approach requires the manual programming of rules to define how the machine operates under certain circumstances by using if-then structures \cite{RuleBased}. However, this manual implementation demanded the development of comprehensive and well-defined rules tailed to the problem at hand, a process known as knowledge engineering \cite{feigenbaum1980knowledge}. With the introduction of the classical machine learning methods in the early 80’s, the hand-designed programming of the rules and the knowledge engineering is replaced by the handcrafted extraction of features \cite{picon2020deep}. The improvement came up with the introduction of machine learning to manage the mapping of handcrafted features using machine intelligence. The next step of this set of improvements was the transfer of the feature extraction task from humans to the intelligent machines \cite{guyon2006introduction}. As the field of deep learning evolved, the feature extraction process became progressive, various and more complex. 
Throughout the machine learning era, many successful feature classifiers have been proposed, and the investigation for new methodologies has become less prominent. However, the pursuit of advancements in feature extraction continues to grow. The introduction of generative models has fundamentally transformed the feature extraction task. In contrast to discriminative classifiers, the primary objective of generative models is to construct a valid joint probability density function for both input and output \cite{harshvardhan2020comprehensive}. In other words, the primary objective of generative models is to develop a specific representation that contains the underlying information of a particular label \cite{abukmeil2021survey}. The Bayesian classifiers can be given as the very first examples of generative models. In Bayesian learning, the decision is made with respect to class conditional probability density function for a particular class, which is a probabilistic model of the label under interest \cite{bielza2014}. With the advancements in deep learning, different generative models that are based on deep neural networks are developed, giving rise to a new family of generative networks known as deep generative models (DGM) \cite{fergus2022deep}. In recent years, generative models have gained popularity for their ability to create realistic data across various domains. Most well-known and widely used generative models are Generative Adversarial Networks (GANs) \cite{goodfellow2014generative}, and Variational Autoencoders (VAEs) \cite{Kingma2014}. The VAEs excel at generating and denoising images, also they are efficient as feature extractors. Another noteworthy generative model is the PixelRNN which can generate high-quality images by processing the spatial information between pixels \cite{van2016pixel}. The Transformer, an architecture that uses self-attention phenomena to model dependencies between the input and output sequences, has also been applied to language generation tasks and has shown remarkable results \cite{vaswani2017attention}. Another significant model is the Generative Flow which is known for its ability to generate high-resolution images and model complex probability distributions \cite{kingma2018glow}. Additionally, ChatGPT is an advanced language model developed by OpenAI, capable of generating coherent and contextually relevant text based on input prompts \cite{Radford2019LanguageMA}. Utilizing the state-of-the-art machine learning techniques, it has success across a variety of natural language processing tasks, making it a valuable tool for diverse applications.  

Discriminative models are widely used in machine learning applications such as classification, regression, and ranking. These models primarily focus on modeling the conditional probability density of classes based on the provided data. As a result, the discriminative models focus on the information that distinguishes between classes. Since the discriminative models are targeting to extract the highest discriminative information, the extracted features are highly dependent with the targeted task. Therefore, the information that is jointly contained in both labels is ignored by the discriminative model, since it is not capable of discriminating the labels. However, discriminative models overlook the shared information present in both labels, as they are not designed to differentiate between them. This approach is efficient and robust for solving various real-life problems. Discriminative models, such as the Support Vector Machine (SVM), Random Forest, and deep learning networks like AlexNet and RNNs, have been widely used in various fields, including classification and natural language processing. These models have demonstrated their effectiveness in pattern recognition and achieving state-of-the-art performance on benchmark datasets \cite{cortes1995support, breiman2001random, krizhevsky2017imagenet, mikolov2010recurrent}. However, we believe that for a model to converge to machine intelligence, it needs to increase its level of perception and understanding of the reality presented to it. Therefore, we claim that the quality of information extracted from the data should be independent of the targeted task. As machine learning is a scientific study, it needs to advance in various directions, including high-quality information retrieval. In summary, while discriminative models have shown great promise and effectiveness in solving many real-world problems, it's crucial to prioritize high-quality information retrieval that is independent of the targeted task. This will help models converge towards machine intelligence and improve their overall understanding and perception of the reality presented to them. 

In the literature, there are several feature selection methods such as NCA and ReliefF. Neighborhood Component Analysis (NCA) is a non-parametric feature selection method that aims to improve the prediction accuracy of regression and classification algorithms. Iterative NCA (INCA) is proposed to overcome optimal feature vector and negative weight issues of NCA. INCA benefits from a loss function to select best features \cite{ASLAN2022104539,liu2022artificial}. Relief based feature selection algorithms are inspired by the original Relief algorithm first developed by Kira and Rendell in 1992. Relief algorithm assesses the efficacy of attributes by analyzing how effectively their values can distinguish instances that are in close proximity to each other. Relief algorithms can only be used for binary classification which involve either discrete or numerical features. ReliefF is a derivative of the original Relief algorithm which solves these drawbacks \cite{robnik2003theoretical}. The Relief feature selection method utilizes Euclidean distance to find near-hit, near-miss instances, while the ReliefF weights are constructed using Manhattan distance \cite{ASLAN2022104539}. One extension of ReliefF algorithm is iterative ReliefF that automatically identifies the most informative features by utilizing a loss calculator to optimize the feature selection process \cite{tuncer2020automated}. There are several studies that are based on either Relief or NCA feature extraction methodologies. Baygin et al. proposed a deep feature generation technique which extract features via DarkNet19 using original CT image and 64 patches of original CT image as inputs. After resulting features are extracted and concatenated, they are selected using the iterative NCA method \cite{BAYGIN2022102274}. Liu et al. proposed a deep feature generation technique which extracts features via 3 best CNNs chosen among 16 CNNs. Original ultrasound image and 8 grids of original image are supplied as input. NCA is used as the first feature selection step after acquired features are merged, iterative NCA is used to decide final features \cite{liu2022artificial}. Aslan et al. proposed a deep feature generation technique which extracts features via 13 different CNNs. Chest X-ray images are used as an input. Feature selection is done in two steps. In the first step iterative NCA is used, and iterative ReliefF is later used to decide the final features \cite{ASLAN2022104539}. Barua et al. proposed a deep feature generation framework which extracts features via 18 different CNNs \cite{e23121651}. Original OCT images and multiple multilevel pooling decompositions of the images are used as input. Extracted features are concatenated and selected using 4 step selection methodology. In the first step, ReliefF is used to select the most informative 1/10th of all features, second and third steps are misclassification rate calculation with SVM and select the best 5 feature vectors among all. At the final step iterative ReliefF is used to decide the final features. 
\begin{figure}[hbt]
    \centering
    \includegraphics[scale=0.1]{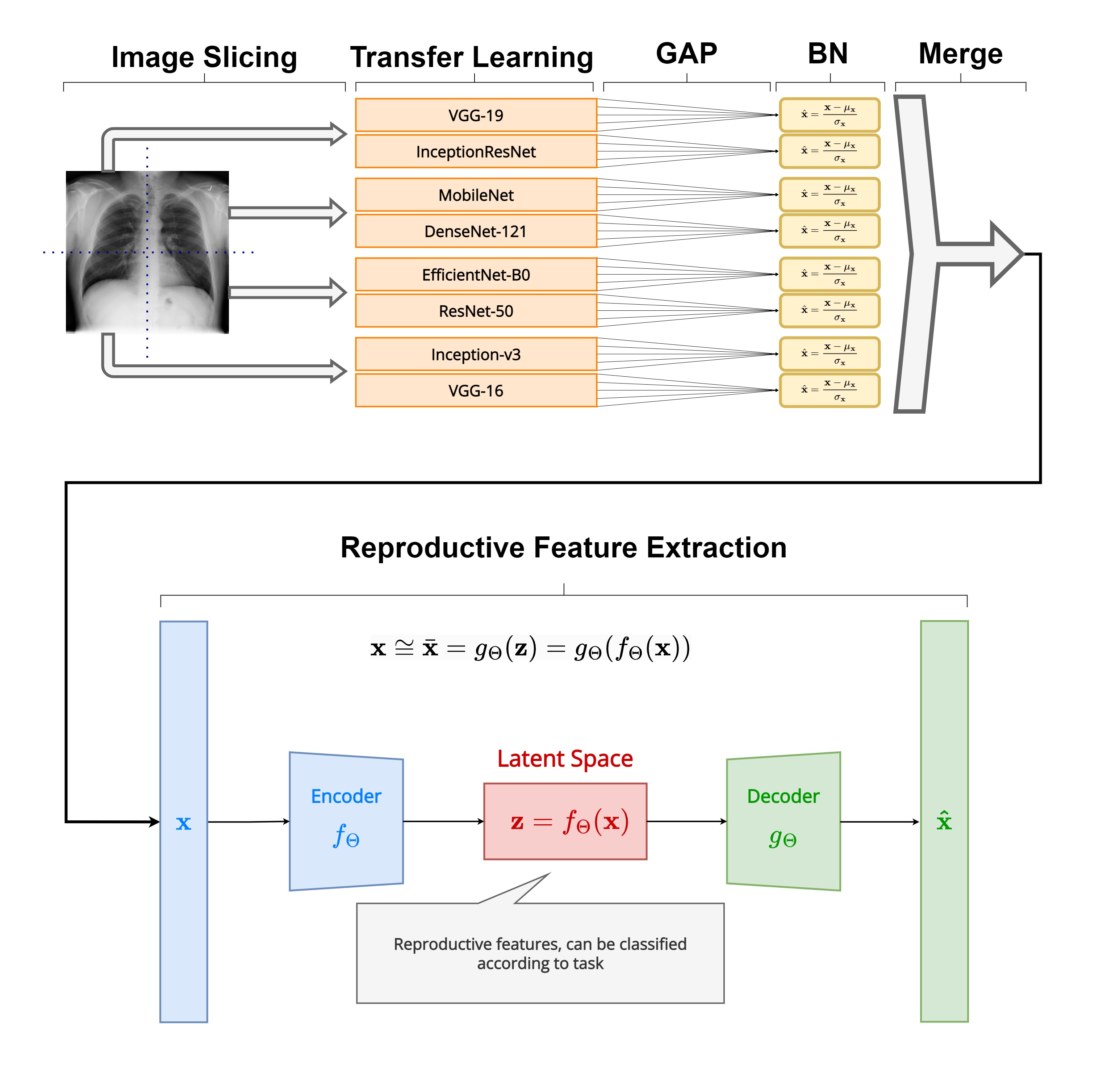}
    \caption{Representation of the proposed deep reproductive feature generation framework.}
\end{figure}

In this paper, we propose a multi-stage feature extraction framework utilizing transfer learning, eight pre-trained deep CNNs, and an autoencoder to diagnose patients' health conditions using chest X-rays. The X-ray images are converted to RGB format and resized to 448x448 resolution before being divided into four equally sized sections. Each section is processed by two state-of-the-art pre-trained deep CNNs with their classifier sections removed. Feature tensors from the CNNs are flattened, concatenated, and normalized. 
To extract high-quality, task-independent, reproductive features, the feature vector is fed to a single-layer autoencoder trained to reproduce the input information. After training, the features obtained from the bottleneck layer are extracted and a latent space that contains reproductive features is obtained. The extracted features from the bottlenech layer are used to classify patients' health conditions. The t-SNE embedding of the latent space is obtained to monitor feature quality.

The key contributions of this study can be summarized as follows: 
\begin{itemize}
    \item Feature extraction from chest x-rays is achieved by using pre-trained deep CNNs. With the help of transfer learning, overfitting, and other difficulties of training CNNs are avoided. 
    \item Employment of eight different pre-trained CNNs allowed the proposed architecture to use different representations of the same information for better feature extraction. 
    \item By using an autoencoder, a reproductive feature extraction is applied to obtain conceptive information from the feature vector obtained with CNNs. 
    \item We developed a framework with affordable computational demands. The training process of the proposed framework is completed in a small time period with an affordable training setup. 
    \item By using different classifiers such as, a shallow multilayer perceptron, a single layer perceptron and a SVM, the discriminative capacity of the reproductive features are observed through the classification accuracy. 
    \item The proposed framework managed to overcome the performances of frameworks which use well-known transfer learning models as feature extractors with the same classifiers. 
    \item Extracting task-independent reproductive features is superior to extracting only task-based discriminative features, as it allows smore flexibility and adaptability to different problems. 
\end{itemize}

\section{Methods}
\subsection{Dataset \& Resources}
\subsubsection{Dataset}
\begin{figure}[H]
    \centering
    \subfigure[]{\includegraphics[scale=0.215]{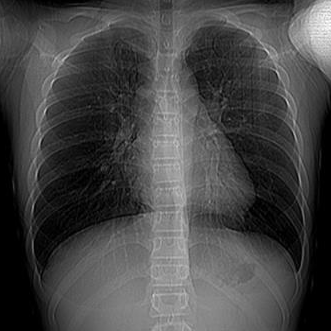}}
    \subfigure[]{\includegraphics[scale=0.07]{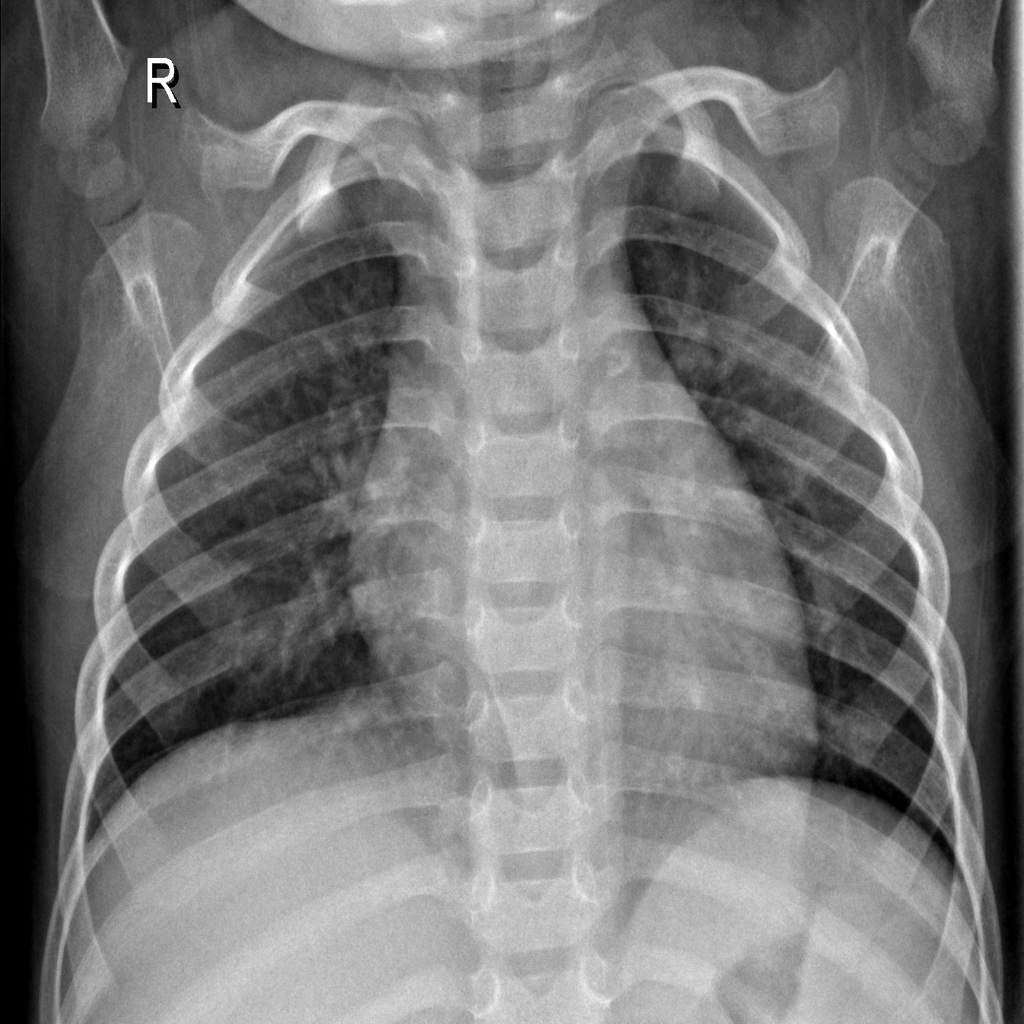}}
    \subfigure[]{\includegraphics[scale=0.07]{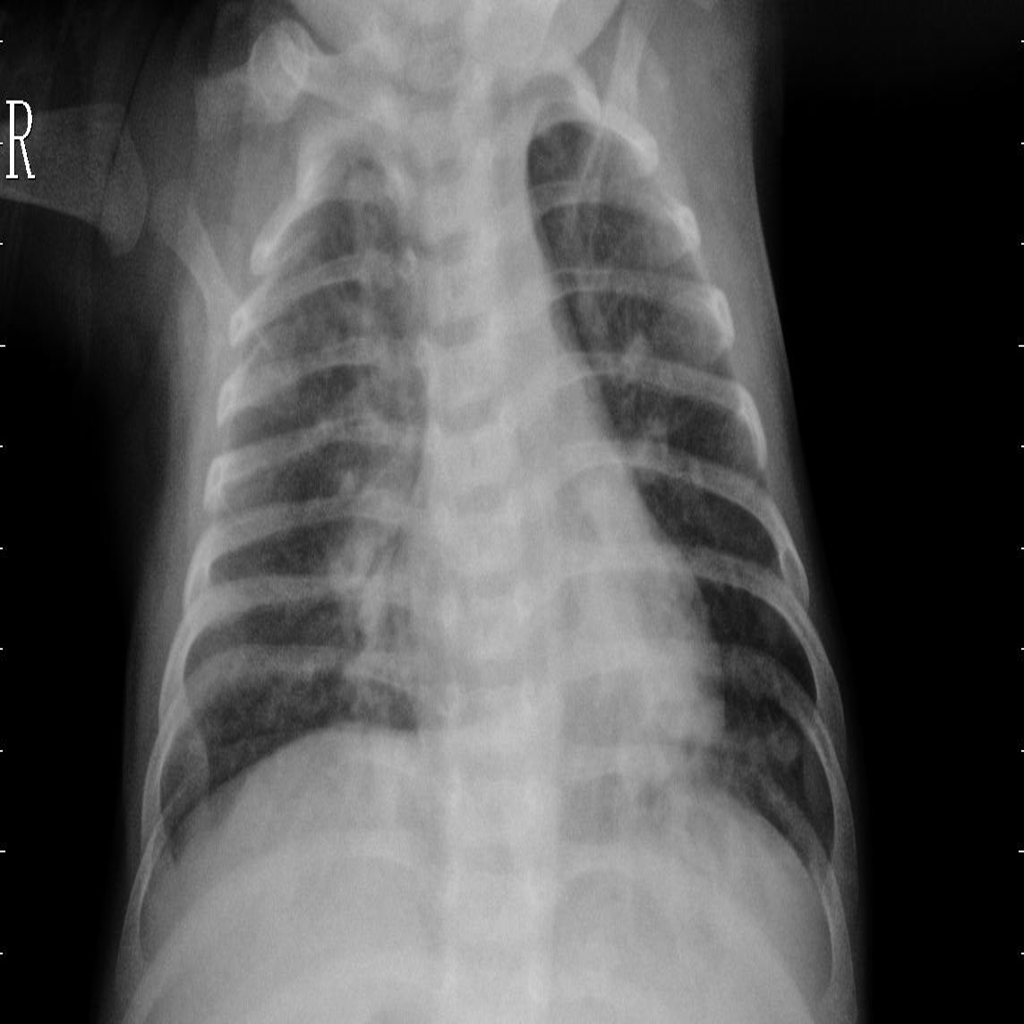}}
    \caption{Example images from the dataset: a) COVID-19, b) Normal, and c) Viral Pnemonia.}
    \label{fig:eximg}
\end{figure}

In this study, we utilized the second version of the COVID-19 Radiography Dataset from Kaggle. A team of researchers from Qatar University, Doha, Qatar, and the University of Dhaka, Bangladesh, along with their collaborators from Pakistan and Malaysia in collaboration with medical doctors have created a database of chest X-ray images for COVID-19 positive cases along with normal and viral pneumonia images. In the second version the dataset contains 1143 chest X-rays with COVID-19 case, 1345 chest X-rays with viral Pneumonia case and 1341 chest X-rays with normal case. The resolution of the X-ray scans was 256x256. Example data samples for each case taken from the dataset is given in Figure \ref{fig:eximg}.

\subsubsection{Data Processing}
The flow of the implemented framework starts with the section wise analysis of an image. Since the proposed algorithm is developed on the basis of transfer learning, the created sections should have a resolution of 224 x 224. Therefore, the chest X-rays are saved as RGB images with a resolution of 448 x 448. Then, the images are converted into three-dimensional tensors for the sectioning operation. Transformation of the RGB images generated a 448x448x3 sized tensor. Finally, by using indexing, the tensor is separated into 4 equally sized (224x224x3) tensors where all contain different parts of the chest X-rays.

\subsubsection{Computational Setup}
Training of the proposed framework and the preprocessing operations are functioned using the Python programming language. We implemented our model in the Keras backend of the TensorFlow 2.11.0 framework. Computational experiments were performed on a desktop with the following properties: Intel i5-12600K CPU, 64 GB RAM, NVIDIA GeForce RTX 3090 Ti 24 GB GPU. The categorical cross-entropy loss function and the Adam optimizer with default settings are used during the training. The batch size and number of epochs are selected as 32 and 40.

\subsection{Architecture of the Proposed Framework}

Convolutional Neural Network (CNN) is a widely used type of deep artificial neural network which has the capability of learning from data as in the image form. They contain two sections: feature extraction and classification. The feature extraction stage of the network consists of stacked convolutional and pooling layers. The classification stage is built by fully connected layers. Convolutional layers extract feature maps that are lessened by pooling layers after applying maximum or averaging operations \cite{lecun1995convolutional}.

The convolutional layer is the elemental layer of the CNNs. It makes feature extraction by applying convolution operations with filters (kernels) to the tensor shaped multi-dimensional data. Each convolutional layer outputs a feature map which is a three-dimensional tensor that stores the results of convolution operations. The following equation describes the behavior of ${n^{th}}$ convolutional layer in a CNN:

\begin{center}
    \centering
    \Large
    $x_i=f\left(\sum_{j=1}^{N}W_k^{n-1}*y_m^{n-1}+b_k^n\right) $
\end{center}

Where ${x_i^n}$ is the ${i^{th}}$ feature map, ${W_k^{(n-1)}}$ and ${b_k^n}$ are the weights of the ${k^{th}}$ filters and the bias term, ${y_m^{(n-1)}}$ is the ${m^{th}}$ feature map in the previous layer, N represents the number of total features and (*) denotes the convolution operation \cite{albawi2017understanding}.

Pooling is an operation that is used for filtering the desired information inside the feature maps in a user-defined method. The pooling layers apply the user-defined operation along the windows sliding over the feature maps. The pooling layers effectively refine the feature maps by decreasing the size and eliminating the undesired information. In this study, we used the max-pooling operation where only the maximum value in the sliding window is selected \cite{lecun1998gradient}. 

\subsection{Transfer Learning}
Training deep CNNs can be challenging due to the need for substantial resources such as large datasets and high-end computation devices since there are major problems to overcome such as overfitting and vanishing gradients. Moreover, they require a precise hyperparameter tuning for obtaining sufficient performances. Transfer learning is a procedure that is based on using a previously trained architecture to avoid these difficulties. In various transfer learning applications, it is proposed to utilize a pre-trained neural network with its weights and modify some parts of the model to train for fine-tuning. Pre-trained networks decrease the required time and resources for the training since they are already trained with intense resources by deep learning specialists \cite{yosinski2014transferable, pan2010survey, lu2015transfer}.

Pre-trained CNNs are networks that are previously trained on massive datasets by deep learning specialists. The VGG-19, EfficientNet-B0, ResNet50, VGG16, DenseNet121, MobileNet, Inception-v3 and InceptionResNet models are used as base models for the feature extraction during the study. Mentioned models are the state-of-the-art pre-trained CNNs that are trained on the ImageNet dataset. ImageNet is a well-known dataset that consists of 14 million images from 20,000 different classes \cite{russakovsky2015imagenet}.  

Researchers have explored the use of pre-trained feature extractors from diverse domains to extract valuable insights from data. Pre-trained feature extractors deliver noteworthy results. As an example, Narin et al. built a feedforward neural network to classify COVID-19 patients based on chest X-Ray pictures and utilized pre-trained CNNs to extract features \cite{narin2021automatic}. In addition, pre-trained CNNs have been applied to extract features from scalograms (combined with SVM classifiers) to detect schizophrenia \cite{shalbaf2020transfer}. Furthermore, pre-trained CNNs have been utilized to extract features from EEG images to identify neonatal seizures \cite{caliskan2021transfer}. Recently, various events recorded by $\Phi$-OTDR sensing system were classified using the pre-trained AlexNet architecture followed by an SVM classifier \cite{li2022quickly}.
\subsubsection{VGG-16 and VGG-19}
VGG is a CNN architecture that is proposed as an improvement over the famous AlexNet architecture \cite{krizhevsky2017imagenet} which is the winner of the ImageNet Large Scale Visual Recognition Challenge (ILSVRC) in 2012. Four different configurations are proposed for the VGG architecture. The VGG16 and VGG19 are the two largest configurations in terms of complexity. The VGG16 and VGG19 architectures consist of five convolutional blocks that have 16 and 19 convolutional layers in total respectively. Both architectures are terminated with a 4-layer feed forward neural network. The total number of parameters in the VGG16 and VGG19 architectures are around 138 and 144 million respectively. Both VGG architectures are also trained on the ImageNet database and became successful \cite{simonyan2015a}. After removing the classifier sections, the VGG architectures yield a feature tensor of shape 7x7x512. 

 \subsubsection{ResNet50, InceptionV3 and InceptionResnetV2}
 Residual Neural Network (ResNet) is a CNN architecture that uses skip connections between layers to improve feature extraction performance. The depth and complexity, causes an information distortion throughout the network. With the help of these shortcuts, ResNet architecture reduces the information distortion. Another tool that ResNet architecture contains to improve the performance are the bottleneck layers which reduce the computational demand of the architecture. The ResNet-50 model contains 50 convolutional layers, has more than 23 million parameters \cite{he2016deep}. The shape of the feature tensor for ResNet50 is 7x7x2048. 
Inception-v3 architecture stands out with its large number of inception blocks. The inception blocks are parallel convolutional layers with different kernel sizes whose outputs are concatenated at the end of each block. With the help of the inception blocks, the number of connections in a model can be significantly removed without making any sacrifices on the performance. Despite having many inception blocks, the Inception-v3 architecture contains approximately 24 million parameters. The Inception-v3 architecture outputs a feature tensor of shape 5x5x2048 \cite{szegedy2016rethinking}. 
The InceptionResNet-v2 is a deep CNN architecture that combines ResNet and Inception models. It contains residual inception blocks which are inception blocks with skip connections. With this fusion implementation, the problem of degradation caused by the depth is solved, additionally the training time is reduced. The InceptionResNet-v2 contains 164 convolutional layers \cite{szegedy2017inception}. After the removal of classifier sections, InceptionResNet-v2’s features are stored in a feature tensor with shape 5x5x1536. 

\subsubsection{MobileNet and EfficientNet-B0}
MobileNet is one of the most famous and efficient CNN architectures that is used in the applications of image classification. The standout point of the MobileNet architecture is its depthwise convolutions that are used instead of standard convolutions. MobileNet is also very useful because of its capability of being tunable. The architecture contains two unique hyperparameters that allows one to decrease the training time and required disk space with price of performance and latency. The MobileNet architecture is small compared to VGG-based models and has approximately 14 million parameters \cite{Kingma2014}. The shape of the feature tensor for MobileNet is 7x7x1024. 
EfficientNet is CNN architecture that is developed with the intention of balancing the architecture's depth, width and resolution. In order to achieve this, the EfficientNet architecture proposes a new scaling technique that uniformly scales the mentioned parameters of the model using a compound coefficient. The new family of models called EfficientNets allowed the reduction of the number of parameters and FLOPs using mobile bottleneck convolutions. Furthermore, the most improved version of the family, the EfficientNetB7 managed to beat other state-of-the-art CNN architectures. In this research, EfficientNetB0, the first member of the EfficientNets family, is used. The EfficientNetB0 contains 9 layers and less than 10 million parameters \cite{tan2019efficientnet}. Before classifying, the EfficientNet flattens its 7x7x1280 shaped feature tensor. 

\subsubsection{DenseNet}
DenseNet-121 is a CNN architecture that is built using dense blocks which are submodules connecting layers. Inside dense blocks, every previous feature map is concatenated into a single tensor and used as an input for the next block. The feature map size is kept the same inside dense blocks. The downsampling operation is used for keeping the feature map size the same and it is performed by transition layers which connect the dense blocks by downsampling the output using convolution and pooling operations. DenseNet-121 architecture consists of 120 convolutional layers however it is one of the lightest CNN architectures with approximately 7.6 million parameters \cite{huang2017densely}. The DenseNet-121 architecture outputs a feature tensor of shape 7x7x1024. 

\subsection{Autoencoder}
Autoencoder is a neural network architecture capable of extracting features from its input in an unsupervised manner. The autoencoder consists of two modules which are encoder and decoder. Although both modules are similar in terms of structure, they have complementary tasks. The encoder applies a non-linear dimensionality reduction operation and creates a more compact representation of the input by extracting valuable features. At the end of this process, the advanced features, extracted from the input, are mapped in the so-called latent space which is the input of the decoder module. The decoder tries to create the inverse of the encoder's operation by using similar non-linear transformations. It tries to reconstruct the input using the features in the latent space. By training both of these modules together, it is aimed to build a model that extracts features that are capable of reconstructing the original data \cite{hinton2006reducing}. Since the objective of the autoencoder is to minimize the difference between the input and the decoder output, its loss function can be written as: 

\begin{center}
\Large
$\mathrm{MSE} = \frac{1}{n} \sum_{i=1}^{n} (x - \hat{x})^2$
\end{center}
where $x$ is the input feature vector, $\hat{x}$ is autoencoders approximation, and $n$ is the dimension of the feature vector.\\

Despite being really powerful feature extractors, autoencoders are highly tending to overfit \cite{sankaran2017group}. Overfitting in an autoencoder occurs when the model learns to perfectly reconstruct the training data but fails to generalize to new, unseen data. This can result in poor performance on new data and reduced ability of the autoencoder to compress the input data into a meaningful lower-dimensional space. In other words, an overfitted autoencoder may have learned to memorize the training data rather than learn the underlying patterns and features of the data. This can lead to poor performance when using the autoencoder for tasks such as data compression or denoising \cite{NEURIPS2020_e33d974a}. To avoid overfitting in autoencoders, several techniques can be used. One common technique is to add noise to the input data during training, which can help the model learn more robust features \cite{steck2020autoencoders}. Another approach is to use regularization techniques such as weight decay or dropout, which can help to prevent the model from overemphasizing any one feature \cite{slatton2014comparison}. However, effects of the mentioned applications, such as adding noise, are highly dependent on the input. If the processed data is already noisy, contains lots of variability or highly sensitive to small changes, adding noise may affect the autoencoder's performance negatively \cite{vincent2010stacked}. 
In the proposed framework, the information fed into the autoencoder is the features extracted with pre-trained CNNs, which is highly sensitive to noise and contains lots of variability. Therefore no noise addition has been done during the training of the autoencoder. 

The latent space of an autoencoder is an abstract representation of the input that extracts the conceptive information of the data. This latent space can be considered as a lower-dimensional representation of the input data, created by a series of nonlinear equations. The structure of the latent space is capable of providing intuitions about the fundamental configuration of the data. The latent space of an autoencoder is a powerful tool for understanding, analyzing, and working with complex, high-dimensional data \cite{Kingma2014}.
\begin{figure}
    \centering
    \includegraphics[scale=0.23]{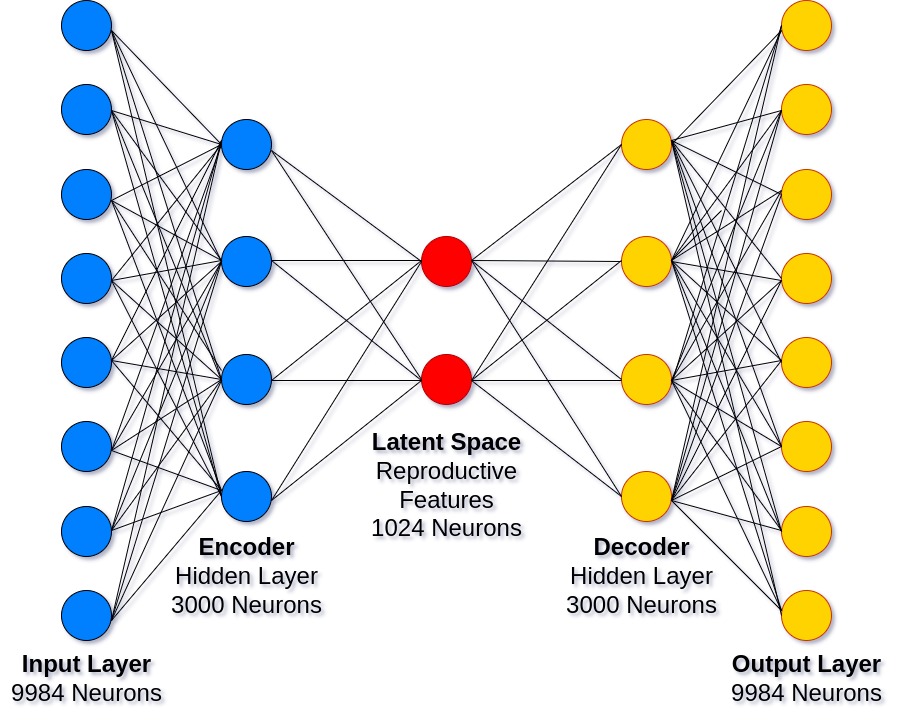}
    \caption{Architecture of the autoencoder used to extract reproductive features.}
    \label{fig:ae}
\end{figure}
\subsection{Classifiers}
\subsubsection{Feedforward Neural Network}
A feedforward neural network is a deep neural network that consists of stacked dense layers. It consists of input, output, and hidden layers. In each layer, the linear combination of previous layer outputs and the layer weights is obtained and fed into a nonlinear activation function. In the majority of the applications, they serve as the classifier section of a CNN since it classifies the flattened features. In this study, the rectified linear unit (ReLU) activation function is used in each hidden layer. In the output layer, the softmax activation function is used to predict the output class of the images. Mathematical descriptions of the mentioned activation functions are given below \cite{lecun2015deep}. 

\begin{center}
\centering
\large
ReLU(x) = $\left\{
        \begin{array}{ll}
            0 & \quad x \leq 0 \\
            x & \quad x \geq 0
        \end{array} \right\}, \quad Softmax(x_{i}) = \frac{e^{x_{i}}}{\sum_{k}{e^{x_{k}}}} $
\end{center}

\subsubsection{SVM}
Support vector machine (SVM) is a supervised learning algorithm that belongs to the linear classifiers family. It simply maps the inputted data to a higher dimensional space to maximize the margin between data points. With this operation, SVM uses hyperplanes and creates linear decision boundaries between the labels \cite{cortes1995support}.
\clearpage
\subsection{Applied Procedure}
The pseudo code representation of the applied procedure is given below.
\begin{algorithm}[H]
\caption{Pseudo Code of the proposed framework}
\begin{algorithmic}[1]
\FOR{trail $\in$ range(50)}
\STATE dataset, labels = \textbf{Load-Dataset}(path)
\STATE
\STATE
\FOR{image $\in$ dataset \& label in labels}
    \STATE img = \textbf{resize}(image, (448,448,3))
    \STATE images.\textbf{append}(img)
    \STATE labels.\textbf{append}(label)
\ENDFOR
\STATE
\STATE images.\textbf{shuffle}(seed)
\STATE labels.\textbf{shuffle}(seed)
\STATE
\FOR{CNN $\in$ CNNs}
    \STATE CNN.\textbf{remove}(classifier)
    \STATE CNN.\textbf{add}(GlobalAveragePooling)
    \STATE CNN.\textbf{add}(BatchNorm)
\ENDFOR
\STATE 
\FOR{each image}
    \STATE sections = \textbf{ImageSlicer}(image)
    \STATE
    \FOR{section $\in$ sections}
        \FOR{$i \in [1,3,5,7]$}
        \STATE $F_{i,i+1}$ = \textbf{concat}[$CNN_i(section_i)$,$CNN_{i+1}(section_i)$]
        \ENDFOR
    \ENDFOR
    \STATE $FV(image)$ = \textbf{concat}[$F_{1,2}$,$F_{3,4}$,$F_{5,6}$,$F_{7,8}$]
\ENDFOR
\STATE 
\STATE X-train, X-test, y-train, y-test = \textbf{train-test-split}(FV, labels)
\STATE Autoencoder.\textbf{fit}(input=X-train, output=X-train, loss=MSE, opt=Adam)
\STATE 
\STATE Latent-space-train = Autoencoder.encoder.\textbf{predict}(X-train)
\STATE Latent-space-test = Autoencoder.encoder.\textbf{predict}(X-test)
\STATE 
\STATE train-embeddings = \textbf{t-SNE}(Latent-space-train)
\STATE test-embeddings = \textbf{t-SNE}(Latent-space-test)
\STATE 
\STATE metrics-SLP = SLP.\textbf{fit}(X=LS-train, y=y-train, loss=CCE, opt=Adam)
\STATE metrics-MLP = MLP.\textbf{fit}(X=LS-train, y=y-train, loss=CCE, opt=Adam)
\STATE metrics-SVM = SVM.\textbf{fit}(X=LS-train, y=y-train)
\STATE 
\STATE metrics.\textbf{append}([metrics-SLP,metrics-MLP,metrics-SVM])
\STATE
\ENDFOR
\end{algorithmic}
\end{algorithm}
\vfill
\section{Results \& Discussion}
\begin{figure}
    \centering
    \includegraphics[scale=0.6]{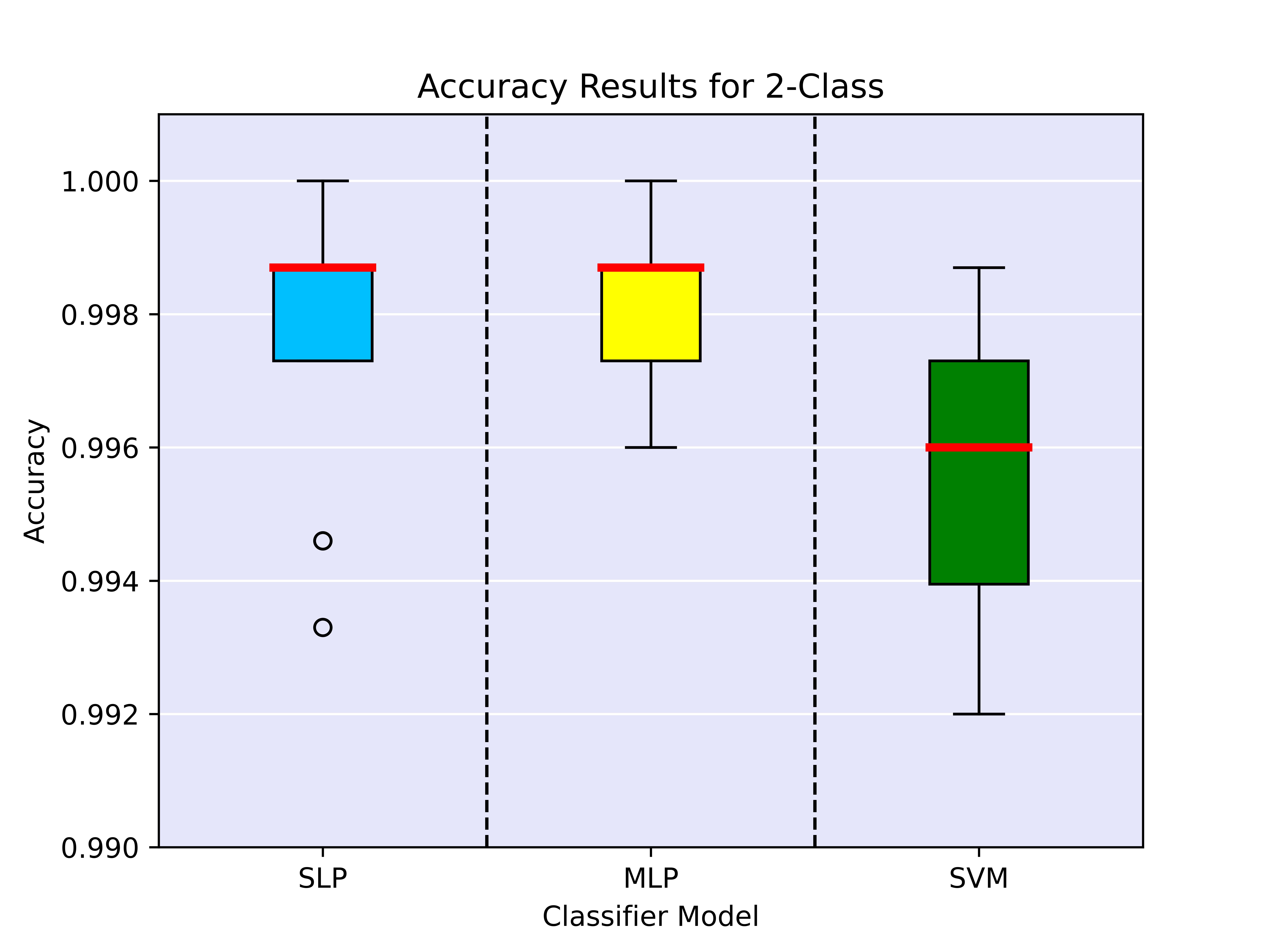}
    \caption{The box plot representation of the average accuracy values after 50 trainings for binary classification.}
    \label{fig:box2}
\end{figure}

\begin{figure}
    \centering
    \includegraphics[scale=0.6]{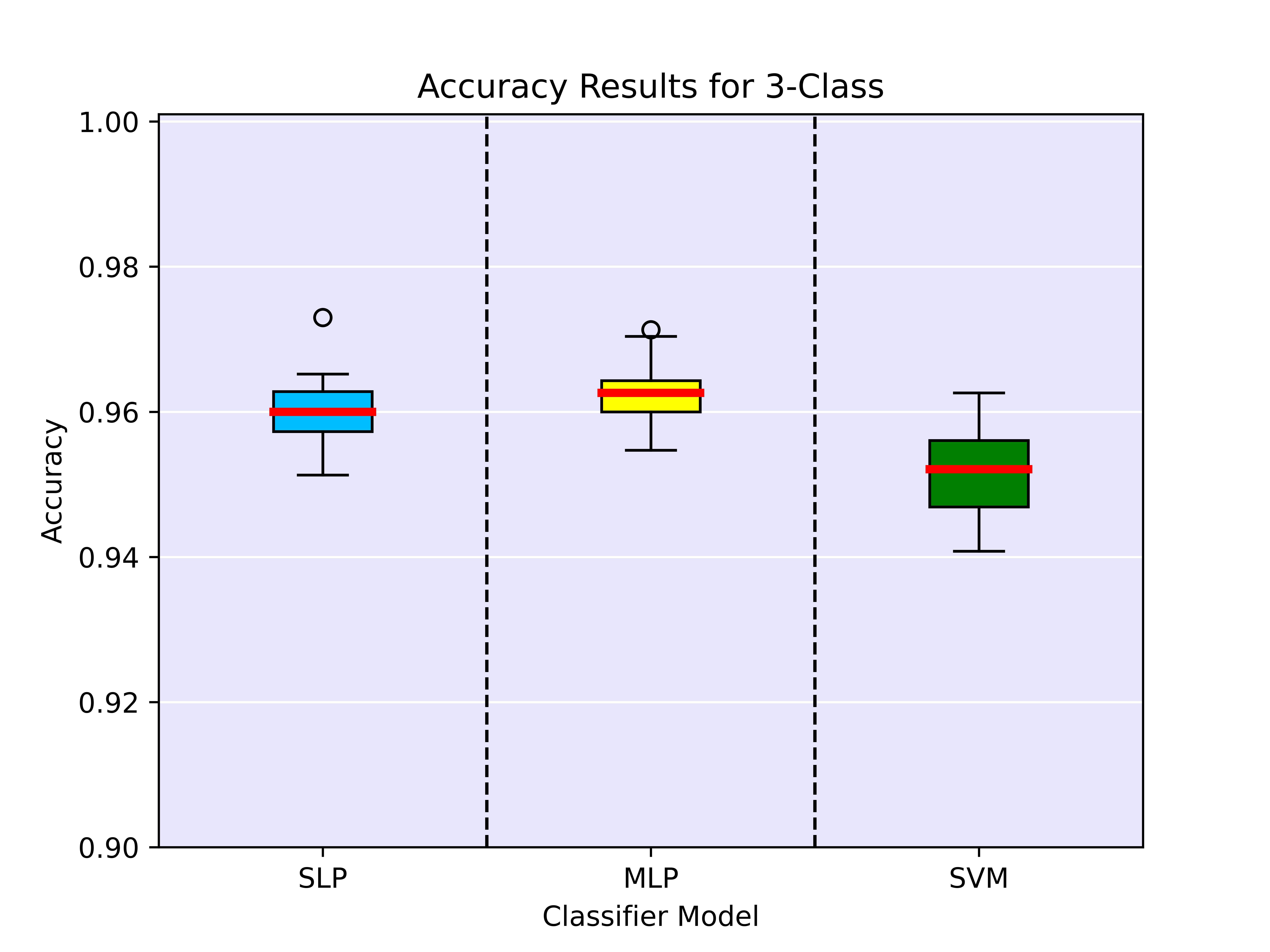}
    \caption{The box plot representation of the average accuracy values after 50 trainings for three-class classification.}
    \label{fig:box3}
\end{figure}

In this research, a multi-stage feature extraction model based on transfer learning and autoencoder is proposed to determine the health condition of patients according to chest X-rays. The proposed methodology achieves an average accuracy of 96.22\% and 99.81\% for three-class and binary classifications, respectively. With the help of transfer learning, the usual difficulties of training deep CNNs, such as large dataset requirements and lack of computational resources, are avoided. In the first feature extraction stage, the image is sliced into four equally sized sections and each section is fed to eight pre-trained CNN architectures whose classifier stages are removed. A global average pooling layer is used to convert output tensors of each CNNs into vectors, which are then concatenated. By applying these operations, different representations of the same information is obtained and represented in the vector form. The CNN architectures employed include VGG-16, VGG-19, ResNet50, DenseNet, MobileNet, EfficientNet, Inception-v3 and InceptionResNet. In the second feature extraction stage, the constructed feature vector is fed to an autoencoder. By training the autoencoder to minimize the difference between encoder input and decoder output, the reproductive features that represent the information more comprehensively are obtained in the bottleneck layer. To assess the model’s performance and the quality of the features, the t-SNE embeddings of the latent space is obtained. Finally, in order to measure the ease of selectivity for the discriminative features, an SVM and a shallow feedforward neural network is used to classify the latent space features. As the box plots show in Figures \ref{fig:box2} and \ref{fig:box3}, both classifiers obtained high classification accuracy values and were capable of selecting the most discriminative features among all of the extracted information. The evaluation of the metrics that represent the training performance for binary classification among "COVID-19" and "Normal" cases are given in Table \ref{tab:2class_metrics}, and the three-class classification results are given in Table \ref{tab:3class_metrics}.

\begin{table}[ht]
\centering
\caption{Average metrics obtained after 50 trainings for binary classification (COVID-19/Normal).}
\begin{adjustbox}{width=1\textwidth,center}
\begin{tabular}{|l||l||cccccccc|c|}
\toprule \toprule
\multirow{3}{*}{Classifier} & \multirow{3}{*}{Metrics} & \multicolumn{9}{c|}{Models} \\
\cmidrule{3-11}
&& VGG-19 & IRN-v2 & MobileNet & EfficientNet & DenseNet & ResNet-50 & Inception-v3 & VGG-16 & \textbf{DRFG} \\
\midrule \midrule
\multirow{4}{*}{Single Layer Perceptron}
& Accuracy & 0.9850 & 0.9931 & 0.9978 & 0.9948 & 0.5805 & 0.9616 & 0.9897 & 0.9890 & \textbf{0.9979}\\
& Precision & 0.9846 & 0.9929 & 0.9977 & 0.9946 & 0.5770 & 0.9610 & 0.9896 & 0.9887 & \textbf{0.9976}\\
& Recall & 0.9853 & 0.9933 & 0.9979 & 0.9950 & 0.5586 & 0.9619 & 0.9897 & 0.9891 & \textbf{0.9975}\\
& F1-Score & 0.9849 & 0.9931 & 0.9978 & 0.9948 & 0.4478 & 0.9613 & 0.9896 & 0.9889 & \textbf{0.9977}\\
\midrule \midrule
\multirow{4}{*}{Multi Layer Perceptron}
& Accuracy & 0.9934 & 0.9936 & 0.9974 & 0.9952 & 0.5394 & 0.9723 & 0.9908 & 0.9961 & \textbf{0.9981}\\
& Precision & 0.9931 & 0.9934 & 0.9973 & 0.9950 & 0.5000 & 0.9727 & 0.9907 & 0.9960 & \textbf{0.9980}\\
& Recall & 0.9936 & 0.9937 & 0.9975 & 0.9953 & 0.2697 & 0.9726 & 0.9907 & 0.9962 & \textbf{0.9981}\\
& F1-Score & 0.9933 & 0.9935 & 0.9974 & 0.9951 & 0.3500 & 0.9721 & 0.9907 & 0.9961 & \textbf{0.9980}\\
\midrule \midrule
\multirow{4}{*}{Support Vector Machine}
& Accuracy & 0.9867 & 0.9895 & 0.9969 & 0.9959 & 0.9839 & 0.9822 & 0.9885  & 0.9900 & \textbf{0.9957}\\
& Precision & 0.9872 & 0.9893 & 0.9968 & 0.9957 & 0.9843 & 0.9827 & 0.9885 & 0.9905 & \textbf{0.9957}\\
& Recall & 0.9862 & 0.9895 & 0.9969 & 0.9960 & 0.9835 & 0.9816 & 0.9884 & 0.9895 & \textbf{0.9957}\\
& F1-Score & 0.9866 & 0.9894 & 0.9968 & 0.9958 & 0.9838 & 0.9821 & 0.9884 & 0.9899 & \textbf{0.9957}\\
\bottomrule \bottomrule
\end{tabular}
\end{adjustbox}
\label{tab:2class_metrics}
\end{table}

\begin{table}[ht]
\centering
\caption{Average metrics obtained after 50 trainings for three-class classification.}
\begin{adjustbox}{width=1\textwidth,center}
\begin{tabular}{|l||l||cccccccc|c|}
\toprule \toprule
\multirow{3}{*}{Classifier} & \multirow{3}{*}{Metrics} & \multicolumn{9}{c|}{Models} \\
\cmidrule{3-11}
&& VGG-19 & IRN-v2 & \textbf{MobileNet} & EfficientNet & DenseNet & ResNet-50 & Inception-v3 & VGG-16 & \textbf{DRFG} \\
\midrule \midrule
\multirow{4}{*}{Single Layer Perceptron}
& Accuracy & 0.9278 & 0.9566 & \textbf{0.9708} & 0.9680 & 0.3738 & 0.8345 & 0.9461 & 0.9411 & \textbf{0.9600}\\
& Precision & 0.9300 & 0.9584 & \textbf{0.9717} & 0.9687 & 0.3639 & 0.8381 & 0.9478 & 0.9425 & \textbf{0.9605}\\
& Recall & 0.9302 & 0.9590 & \textbf{0.9722} & 0.9697 & 0.2218 & 0.8371 & 0.9483 & 0.9428 & \textbf{0.9608}\\
& F1-Score & 0.9298 & 0.9580 & \textbf{0.9718} & 0.9690 & 0.2114 & 0.8339 & 0.9477 & 0.9425 & \textbf{0.9605}\\
\midrule \midrule
\multirow{4}{*}{Multi Layer Perceptron}
& Accuracy & 0.9528 & 0.9533 & \textbf{0.9706} & 0.9685 & 0.3449 & 0.8880 & 0.9420 & 0.9615 & \textbf{0.9622}\\
& Precision & 0.9538 & 0.9540 & \textbf{0.9715} & 0.9695 & 0.3333 & 0.8887 & 0.9440 & 0.9627 & \textbf{0.9633}\\
& Recall & 0.9560 & 0.9563 & \textbf{0.9718} & 0.9700 & 0.1149 & 0.9006 & 0.9445 & 0.9639 & \textbf{0.9633}\\
& F1-Score & 0.9540 & 0.9548 & \textbf{0.9716} & 0.9696 & 0.1709 & 0.8886 & 0.9438 & 0.9628 & \textbf{0.9632}\\
\midrule \midrule
\multirow{4}{*}{Support Vector Machine}
& Accuracy & 0.9474 & 0.9895 & \textbf{0.9722} & 0.9724 & 0.9267 & 0.9348 & 0.9521  & 0.9519 & \textbf{0.9535}\\
& Precision & 0.9497 & 0.9893 & \textbf{0.9730} & 0.9731 & 0.9280 & 0.9364 & 0.9539 & 0.9538 & \textbf{0.9624}\\
& Recall & 0.9481 & 0.9895 & \textbf{0.9732} & 0.9739 & 0.9289 & 0.9355 & 0.9536 & 0.9521 & \textbf{0.9627}\\
& F1-Score & 0.9484 & 0.9894 & \textbf{0.9730} & 0.9734 & 0.9281 & 0.9353 & 0.9535 & 0.9527 & \textbf{0.9623}\\
\bottomrule \bottomrule
\end{tabular}
\end{adjustbox}
\label{tab:3class_metrics}
\end{table}

The coronavirus disease 2019 (COVID-19) has been the most significant public health issue for the past four years. Due to the pandemic, pneumonia-related deaths have rapidly increased making early detection a critical issue that also lacks automation. Chest radiography is a widely used method for diagnosing pneumonia cases due to its cost-effectiveness and speed. However expert support is still required to make diagnostic decisions from X-rays, which limits its accessibility. Nevertheless, the field of deep learning has shown promising results in the analysis of medical image data, and the number of studies on the application of deep learning methodologies in the medical domain are increasing rapidly. Deep learning has shown great potential in the field of biomedical research and has been applied in different domains, such as drug discovery, medical imaging, and genomics. In a study by Alipanahi et al., a deep learning model called DeepBind was developed to predict protein-DNA interactions using a large dataset of DNA sequences and protein binding preferences \cite{alipanahi2015predicting}. DeepBind was able to outperform existing methods. In another study by Litjens et al., deep learning was applied to the task of breast cancer histology image analysis, where a convolutional neural network (CNN) was trained to identify cancerous regions in whole-slide images \cite{litjens2017survey}. Their CNN achieved an accuracy of 81\%, performing better than classical machine learning methods. In a different study, DeepVariant, a deep learning model for variant calling, was developed by Poplin et al. and applied to whole-genome sequencing data \cite{poplin2017scaling}. DeepVariant achieved higher accuracy compared to existing methods. 
However, the uniqueness and vastness of biological data require a more concentrated feature extraction methodology to achieve remarkable outcomes.  Comprehensive analysis of the biological data should capture all available information, as the potential for un-discovered relations in the data is immense. For this reason, a framework employing an unsupervised method to extract reproductive features from the data is crucial. Such a framework allows for the extraction and analysis of conceptive information, which can then be applied to various tasks and investigations.
The main goal of this study is to construct a framework that is based on reproductive information extraction from the data. It is aimed to demonstrate that the selection of the discriminative features from the reproductive features is a valid technique for solving image classification tasks. This study also differentiates from the previous research in the literature in terms of methodology since the different segments of images are analyzed using different pre-trained deep CNNs. 

\begin{figure}
    \centering
    \includegraphics[scale=0.58]{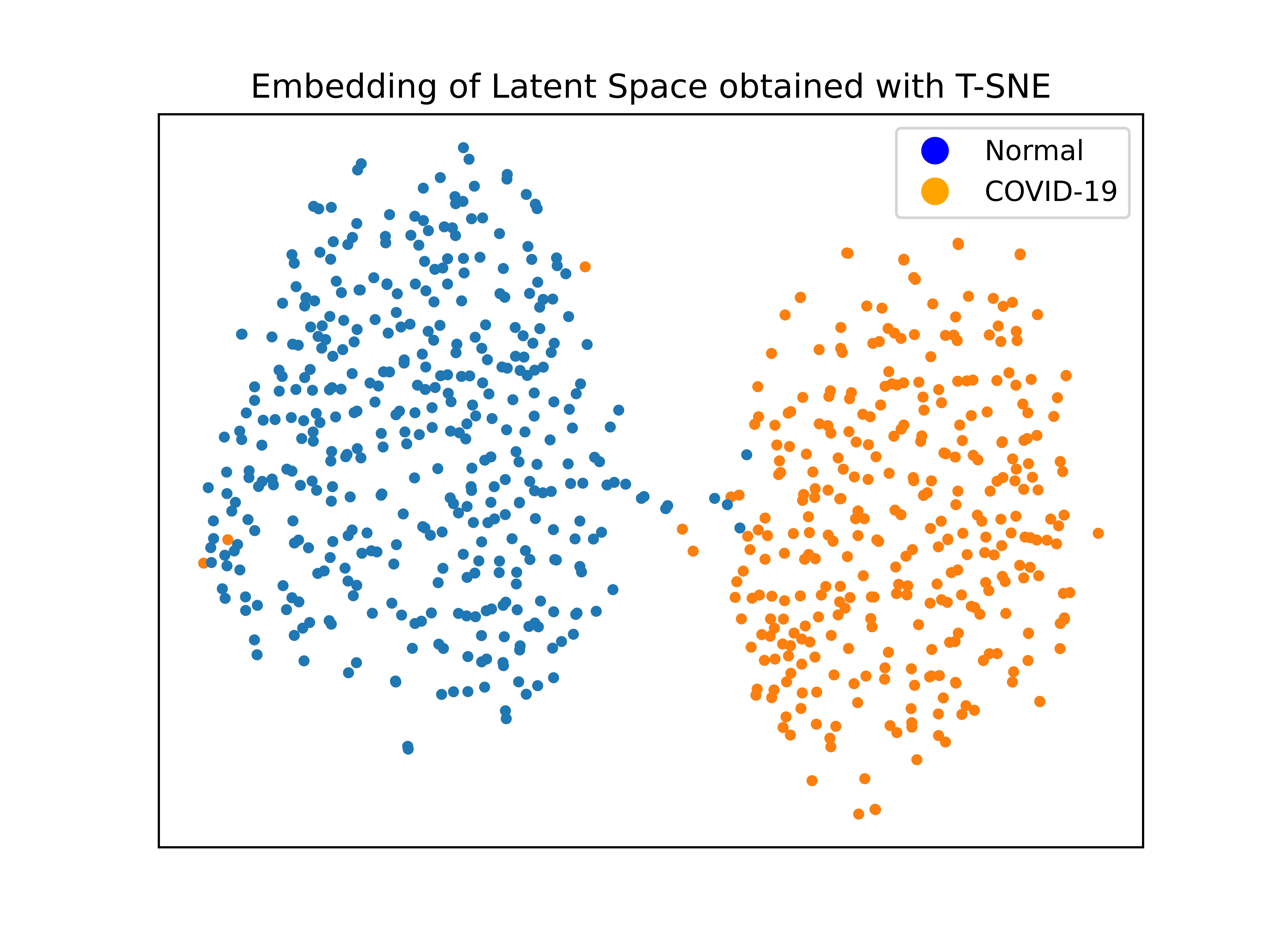}
         \caption{t-SNE embedding of the extracted deep reproductive features during binary classification.}
    \label{fig:embed_2}
\end{figure}

\begin{figure}
    \centering
    \includegraphics[scale=0.58]{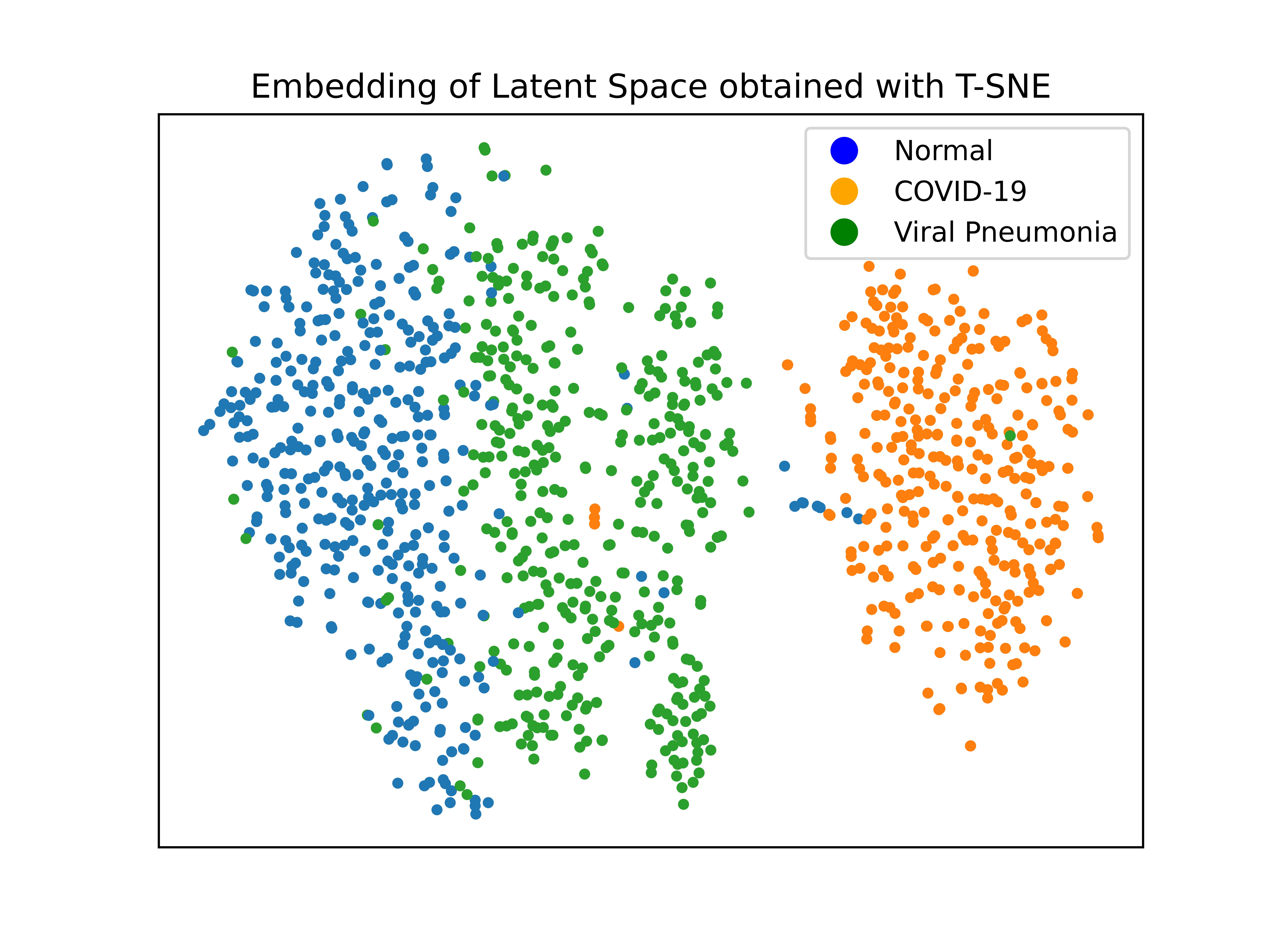}
    \caption{t-SNE embedding of the extracted deep reproductive features during three-class classification.}
    \label{fig:embed_3}
\end{figure}

Furthermore, the extracted information is processed in an autoencoder with the intention of obtaining reproductive features. As Figure \ref{fig:embed_2} and Figure \ref{fig:embed_3} show, the proposed two stage feature extraction methodology, provided the extraction of high quality features. The extracted methodology is novel in the field of deep learning applications on medical studies. 
 The crucial points of this research study can be expressed as: (i) Health condition of patients is determined using the chest x-ray scans, (ii) X-ray scans are sliced into four equally sized square sections, (iii) analysis of each section is completed using two deep pre-trained CNNs, (iv) an autoencoder is used to extract reproductive features from the information obtained from eight different CNNs. (v) by selecting the most discriminative features among the reproductive features, high classification accuracy is obtained, (vi) a new framework based on 2-stage feature extraction is proposed.
 

In order to monitor the performance of the proposed feature extraction framework, the well-known transfer learning models are used as benchmark feature extractors. By using the same classifiers, the feature extraction performance of the proposed framework is compared with the well-known pre-trained transfer learning models. 
Table \ref{tab:2class_metrics} shows that the proposed feature extraction framework achieved a classification accuracy of 99.81\% in binary classification of X-ray scans as COVID-19 and non-COVID-19, outperforming other frameworks with pre-trained feature extractors. Examining the three-class classification results in Table \ref{tab:3class_metrics}, it can be seen that the proposed framework is surpassed only by the frameworks that use MobileNet and EfficientNet as feature extractors, with a classification accuracy difference of less than 1\%. This result indicates that the internal architectures of these two CNNs are particularly suitable for the X-ray data. On the other hand, features extracted with DenseNet architecture require a more sophisticated classifier network to achieve satisfactory accuracy, leading to the conclusion that the DenseNet architecture is not as suitable as MobileNet and EfficientNet. 
These results highlight that the quality and the suitability of the extracted features for discriminative models are highly dependent on the task. Therefore, extracting discriminative features requires a task-oriented and detailed implementation. However, since the proposed feature extraction framework combines and analyzes information obtained from many different feature extractors with a reproductive approach, it is task-independent and suitable for addressing various problems. 
In addition to MLP and SLP classifiers, an SVM classifier is also employed to evaluate the performance of the implemented framework. The main reason behind this procedure is that, instead of compressing reproductive features into smaller spaces and finding non-linear relations, the SVM classifier projects the data into a higher dimensional space and classifies it using linear hyperplanes. Since the approach of SVM is based on projecting into higher dimensional space, its approach is fundamentally different from the MLP and SLP classifiers. As Tables \ref{tab:2class_metrics} and \ref{tab:3class_metrics} show, the proposed framework performs well with all mentioned classifiers, regardless of the approaches. The classification accuracies obtained with SVMs indicate that the extracted features also provide valuable information in the higher dimensional spaces. 
To closely monitor the performance of our feature extraction methodology, feature embeddings are generated by applying t-distributed stochastic neighbor embedding (t-SNE) for the features obtained from the latent space. t-SNE is a widely used dimensionality reduction technique for visualizing high-dimensional data by projecting it onto a lower dimensional plane. The positions of data points in this new plane are determined by their pairwise distances in the original data space. Consequently, while well-organized data may not necessarily result in a satisfactory representation in the 2D space, a well-represented visualization in the 2D space certainly indicates well-organized data \cite{van2008visualizing}. The embeddings of the reproductive features obtained from the latent space are given in Figure \ref{fig:embed_2} and Figure \ref{fig:embed_3} for binary and three-class classification cases. In both embeddings, it can be seen that the proposed feature extraction framework managed to cluster the samples according to their respective classes with clear distinctions. Another notable observation is that, in the three class embedding, the “COVID-19“ and “Normal” classes clusters are considerably separate from each other, while the “Viral Pneumonia” cluster is situated between them. Given the inherent difficulty in differentiating viral pneumonia cases from COVID-19 cases, the proximity of the “Viral Pneumonia” cluster to the “COVID-19” cluster further indicates that the latent space features are not only meaningful but also reflective of real-world relationships.

The implemented two-stage feature extraction methodology is based on producing reproductive features instead of discriminative features. In the proposed methodology the feature extraction operation is done by transfer learning and autoencoder. The classifier module that terminates the framework is trained with the goal of classifying the extracted features, therefore it needs to extract the discriminative information from its input. Given the high classification accuracy, it can be inferred that the classifier modules can effectively select the most discriminative features among the reproductive features. Another conclusion drawn from these results is that the discriminative features is a subset of the reproductive features obtained by the two-stage feature extraction process. This finding suggests that extracting task-independent features capable of recreating the input is superior to extracting only task-based features. The main reason for this superiority is related to efficiency. The resources required to extract reproductive or discriminative features are similar, as both tasks demand large amounts of data and high-end computational devices. However, selecting the discriminative (task-based) features from the reproductive features is a simpler task, as demonstrated by our shallow classifier modules. Additionally, the optimal approach for extracting discriminative features may vary depending on the task. For instance, features required for discriminating chest X-ray scans based on health conditions and smoking habits may share similarities, so the same kind of framework may be useful. In contrast, discriminating chest X-ray scans with respect to patients’ gender requires a different set of features therefore a different feature extraction methodology. On the other hand, the reproductive features extracted from the x-ray scans only require training of a small classifier module that selects the appropriate features for the targeted task. As an example, the latent space obtained from a dataset can be utilized for adressing different problems. The authors claim that the flexibility and task-independence of the reproductive features make the conceptive information approach more favorable.
Deep learning based methodologies hold great promise for making discoveries in various domains. To achieve this, models that are capable of extracting conceptive information from data are essential. Numerous tools in deep learning field are suitable for creating generative models. Moreover, operations performed without machine learning’s involvement, such as image slicing in this application, can be altered to different operations for creating diversity. By focusing reproductive features and their flexibility, researchers can develop more adaptable and efficient models to tackle a wide range of problems in various domains. We believe that the proposed methodology can be further improved and extended to other medical imaging domains, enabling more effective and accurate diagnosis of diseases.

\section{Conclusion}
In this research, we have proposed and implemented a novel multi-stage feature extraction framework for determining patients' health condition using chest X-ray images. The framework combines the power of transfer learning and autoencoder, which results in high-quality reproductive features. The proposed two-stage feature extraction framework differentiates itself from the existing methodologies by dividing each X-ray scan into four equally sized sections and processing them using multiple pre-trained deep CNNs modified by the removal of the classifier modules and addition of global average pooling and batch normalization layers. Feature vectors obtained from the CNNs are concatenated to form a single 9984-length feature vector representing a single X-ray scan. Subsequently, an autoencoder with three hidden layers is trained to extract reproductive features from these vectors, which are then obtained from the latent space of the autoencoder. By training a shallow classifier module, the health condition of the patients is classified. Our methodology demonstrates its effectiveness by achieving high classification accuracies of 96.22\% and 99.81\% for three-class and binary classifications, respectively. The high classification accuracy indicates that the discriminative features can be effectively selected from the reproductive features, making the framework task-independent and suitable for addressing various problems. Our study demonstrates that focusing on reproductive features and their flexibility can lead to the development of more adaptable and efficient models for tackling a wide range of problems in different domains. The results also suggest that extracting task-independent features capable of recreating the input is superior to extracting only task-based features, as it allows for more efficient feature selection. Our approach is novel and shows promising results for analyzing medical image data. In future research, we plan to investigate the applicability of our proposed framework to other medical imaging modalities and different medical conditions. We also intend to explore the integration of additional feature extraction techniques and classifiers to enhance the performance and the adaptability of our framework further. By building upon the foundation of this study, we hope to contribute to the advancement of deep learning applications in medical studies and facilitate more efficient and accurate diagnosis and treatment of various health conditions.

\bibliographystyle{unsrt}  
\bibliography{bibtext_refs}

\end{document}